\documentclass[11pt,a4paper,reqno]{article}
\usepackage{mysty}
\usepackage{authblk}

\usepackage{fullpage}
\usepackage{subfig}

\usepackage{amsfonts,amssymb}
\usepackage{amsmath}
\usepackage{graphicx}
\usepackage{cite}
\usepackage{enumerate}

\SetLabelAlign{parright}{\parbox[t]{\labelwidth}{\raggedleft{#1}}}
\setlist[description]{style=multiline,topsep=4pt,align=parright}

\makeatletter
\let\reftagform@=\tagform@
\def\tagform@#1{\maketag@@@{(\ignorespaces\textcolor{black}{#1}\unskip\@@italiccorr)}}
\newcommand{\iref}[1]{\textup{\reftagform@{\tcr{\ref{#1}}}}}
\makeatother

\begin{document}
	
\title{Design Efficient Exponential Time Differencing method For Hodgkin-Huxley Neural Networks} 
\author{Zhong-Qi Kyle Tian and Douglas Zhou\textsuperscript{\footnote{zdz@sjtu.edu.cn}}}
\affil{School of Mathematical Sciences, MOE-LSC, and Institute of Natural Sciences, Shanghai Jiao Tong University, Shanghai, China}

\date{}
\maketitle

\begin{abstract}
	The  exponential time differencing (ETD) method allows using a large time step to efficiently evolve the stiff system such as Hodgkin-Huxley (HH) neural networks. 
	For pulse-coupled HH networks, the synaptic spike times 
	cannot be predetermined and are convoluted with neuron's trajectory itself.
	This presents a challenging issue for the design of an efficient numerical simulation algorithm.
	The stiffness in the HH equations are quite different between the spike and non-spike regions.  Here, we design a second-order adaptive exponential time differencing algorithm (AETD2) for the numerical evolution of HH neural networks.  
	Compared with the regular second-order Runge-Kutta method (RK2), our AETD2 method can use time steps one order of magnitude larger and improve computational efficiency more than ten times while excellently capturing accurate traces of membrane potentials of
	HH neurons. This high accuracy and efficiency can be robustly obtained and do not depend on
	the dynamical regimes, connectivity structure
	or the network size. 
\end{abstract}

\section{Introduction}

The Hodgkin-Huxley (HH) model \cite{hodgkin1952quantitative,hassard1978bifurcation,dayan2003theoretical}
is a classical neuron model, originally proposed to describe the behaviors
of action potentials of the squid's giant axon. It provides a useful
mechanism that accounts for the detailed generation of action potentials
and the existence of the absolute refractory periods. It also serves
as the foundation for other neuron models such as the one that can
describe the behaviors of bursting and adaption \cite{pospischil2008minimal}.
However, the HH equations are so complicated that it is difficult
to study its properties analytically such as the Hopf bifurcation and
chaotic dynamics \cite{aihara1986chaotic,hansel1996chaos,guckenheimer2002chaos,lin2006entrainment}.
Therefore, its investigation often relies on numerical simulations,
for example, by the Runge-Kutta (RK) methods. 

There are several difficulties to design an efficient and accurate numerical
algorithm for the HH neural network, especially when the network size is large. First,
when an HH neuron driven by external input generates an action potential (the interval of action potential is called spike period in this work), the HH neuron equations become stiff. Regular
RK methods have to use very small time step
to satisfy the requirement of numerical stability \cite{guckenheimer2002chaos,borgers2005background,kassam2005fourth,borgers2013exponential}.
This small time step will significantly increase the computational cost
when studying long time behavior of large-scale HH networks such as 
chaotic attractor dynamics or collecting reliable statistical information of HH neurons such as the distribution of inter-spike-intervals.

The exponential time differencing (ETD) method \cite{hochbruck1998exponential,cox2002exponential,kassam2005fourth,de2008exponential,nie2008compact,hochbruck2010exponential} is proposed for efficient simulation of stiff ordinary differential equations (ODEs). 
The basic idea is to decompose the ODEs into a linear stiff part and a nonlinear non-stiff part. Then, the linear stiff part can be solved by using the integrating factor method, while the nonlinear non-stiff part can be approximated by numerical quadrature.
A second-order ETD method for HH neural networks has been proposed in a recent work \cite{borgers2013exponential}, 
which allows using a large time step to raise computational efficiency. 
However, the HH neural network considered in previous works is a special case where 
each neuron is driven by constant input and the synaptic conductance is described by a smooth function \cite{ermentrout1998fine}.  
For realistic situations, the neurons are generally driven by stochastic spike input and the interaction term is usually
modeled by a Dirac delta function (pulse-coupled), while the
spike-induced conductance dynamics are modeled by
an $\alpha$ function \cite{Somers1995,Hansel1998,sun2009library}.
These make the system become non-smooth and event-driven, while providing challenges for the design of efficient numerical simulation algorithms.
For instance, it is impossible to predetermine the synaptic spike times since they are convoluted with neurons' trajectories themselves.
As a result, one has to evolve the HH network by ignoring the spike interactions among neurons and then use spike-spike interaction to amend the neurons' trajectories at the end of the time
step \cite{brette2007simulation,Hansel1998}. Without a careful recalibration
for the neuronal spikes, the numerical algorithm often suffers from the issue of instability or relatively low numerical accuracy. 

In this work, we first provide a second-order ETD method (ETD2) to evolve a pulse-coupled HH neural network driven by stochastic spike input. 
Note that the stiffness of HH equations are quite different between the spike and non-spike periods, and we find that the ETD method may introduce a relatively large error in the membrane potentials in the non-spike period if using the same time step as that in the spike period.
We then design an adaptive ETD2 method (AETD2) that using different decompositions of the linear and nonlinear parts in spike and non-spike periods. In addition, for the situation where neurons generate spikes in the time step, the effects of the spikes are carefully recalibrated in our AETD2 method to achieve a second-order numerical accuracy. Our AETD2 method is capable of using a large time step, while achieving the same high accurate traces of membrane potentials of each neuron as the RK2 method using a very small time step.  It can improve computational efficiency more than one order of magnitude compared with the RK2 method. This high numerical accuracy and computational efficiency can be achieved over a wide range of dynamical regimes and does not depend on the network connectivity and size.


\section{MATERIALS AND METHODS}

\subsection{The model \label{sec:The-model}}

The dynamics of the $i$th neuron of an HH neural network
is governed by 

\begin{equation}
C\frac{dV_{i}}{dt}=-(V_{i}-V_{Na})G_{Na}m_{i}^{3}h_{i}-(V_{i}-V_{K})G_{K}n_{i}^{4}-(V_{i}-V_{L})G_{L}+I_{i}^{\textrm{input}},\label{eq: V of HH}
\end{equation}

\begin{equation}
\frac{dz_{i}}{dt}=(1-z_{i})\alpha_{z}(V_{i})-z_{i}\beta_{z}(V_{i}), \,\,\,  \text{ for }z=m,h,n,\label{eq:mhn of HH}
\end{equation}
where $C$ is the cell membrane capacitance, $V_{i}$ is the membrane
potential, $m_{i}$, $h_{i}$ , and $n_{i}$ are gating variables for
sodium and potassium currents, respectively \cite{dayan2001theoretical}. The parameters
$V_{Na},V_{K}$, and $V_{L}$ are the reversal potentials for the sodium,
potassium, and leak currents, respectively, $G_{Na},G_{K}$ and $G_{L}$
are the corresponding maximum conductances. The form of $\alpha_{z}$
and $\beta_{z}$ are set as \cite{dayan2001theoretical}:
$\alpha_{m}(V)=(0.1V+4)/(1-\exp(-0.1V-4))$,
$\beta_{m}(V)=4\exp(-(V+65)/18)$, $\alpha_{h}(V)=0.07\exp(-(V+65)/20)$,
$\beta_{h}(V)=1/(1+\exp(-3.5-0.1V))$, $\alpha_{n}(V)=(0.01V+0.55)/(1-\exp(-0.1V-5.5))$,
and $\beta_{n}(V)=0.125\exp(-(V+65)/80)$.

The input current $I_{i}^{\textrm{input}}$ is given
by 
\begin{equation}
I_{i}^{\textrm{input}}=-G_{i}^{E}(t)(V_{i}-V_{G}^{E})-G_{i}^{I}(t)(V_{i}-V_{G}^{I}),
\end{equation}
where $G_{i}^{E}$ and $G_{i}^{I}$ are excitatory and inhibitory
conductances, respectively, $V_{G}^{E}$ and $V_{G}^{I}$ are the corresponding
reversal potentials. The dynamics of conductances can be explicitly
expressed as 

\begin{equation}
G_{i}^{E}(t)=f\sum_{l}H(\sigma_{d}^{E},\sigma_{r}^{E},t-s_{il})+\sum_{j}S_{ij}^{E}\sum_{l}H(\sigma_{d}^{E},\sigma_{r}^{E},t-\tau_{jl}),\label{eq:GE}
\end{equation}

\begin{equation}
G_{i}^{I}(t)=\sum_{j}S_{ij}^{I}\sum_{l}H(\sigma_{d}^{I},\sigma_{r}^{I},t-\tau_{jl}),\label{eq:GI}
\end{equation}
where $s_{il}$ is the spike time of the feedforward Poisson input
with strength $f$ and rate $\nu$, and $\tau_{jl}$ is the $l$th
spike time of the $j$th neuron. The spike-induced conductance
change is described as 

\begin{equation}
H(\sigma_{d},\sigma_{r},t)=\frac{\sigma_{d}\sigma_{r}}{\sigma_{d}-\sigma_{r}}(e^{-t/\sigma_{d}}-e^{-t/\sigma_{r}})\Theta(t),\label{eq:conductance form}
\end{equation}
where $\sigma_{d}$ and $\sigma_{r}$ are slow decay and fast rise
time scale, respectively, and $\Theta(\cdot)$ is the Heaviside function.
Each neuron is either excitatory or inhibitory and its coupling strength
is labeled by its type $E$ or $I$, respectively. For example, $S_{ij}^{E}$
($S_{ij}^{I}$) is the coupling strength from the $j$th excitatory
(inhibitory) neuron to its postsynaptic $i$th neuron. The model parameters
are $C=1\mu\textrm{F\ensuremath{\cdot}cm}^{-2}$, $V_{Na}=50$ mV,
$V_{K}=-77$ mV, $V_{L}=-54.387$ mV, $G_{Na}=120\textrm{ mS\ensuremath{\cdot}cm}^{-2}$,
$G_{K}=36\textrm{ mS\ensuremath{\cdot}cm}^{-2}$, $G_{L}=0.3\textrm{ mS\ensuremath{\cdot}cm}^{-2}$,
$V_{G}^{E}=0$ mV, $V_{G}^{I}=-80$ mV, $\sigma_{r}^{E}=0.5$ ms,
$\sigma_{d}^{E}=3.0$ ms, $\sigma_{r}^{I}=0.5$ ms, and $\sigma_{d}^{I}=7.0$
ms \cite{dayan2001theoretical}. 

The voltage $V_{i}$ evolves continuously according to Equations 
(\ref{eq: V of HH}) and (\ref{eq:mhn of HH}). When it reaches the threshold
$V^{\textrm{th}}=-50$ mV, we say the $i$th neuron generates a 
spike at this time, say $\tau_{il}$. Then it will trigger its postsynaptic
$j$th neuron's conductance change in the
form of $S_{ji}^{Q}H(\sigma_{d}^{Q},\sigma_{r}^{Q},t-\tau_{il})$,
$Q=E,I$.  For the ease of discussion about our algorithm design, we consider an all-to-all connected
network with $S_{ij}^{Q}=S/N$, where $Q=E,I$, $S$ is the coupling
strength and $N$ is the total number of neurons in the network.
Note that our algorithm can be easily extended to networks with more complicated connectivity structure.

\subsection{Runge-Kutta method}
Without loss of generality, we consider the second-order Runge-Kutta method (RK2) as the benchmark and compare it with the ETD methods.
We first introduce the RK2 method to evolve the HH neural network with a fixed time step $\Delta t$, for example, to evolve the system from time $t=t_k=k\Delta t$ to $t=t_{k+1}=(k+1)\Delta t$. 
Since the synaptic spike times in $[t_k, t_{k+1}]$ can not be predetermined, 
one has to evolve the network without considering synaptic spike interactions and reconsider their effects by using spike-spike interactions at the 
end of time step \cite{brette2007simulation,Hansel1998}. 

Due to the pulse-coupled dynamics in Equation (\ref{eq:conductance form}), the numerical accuracy may be very low if the spike timing is not well estimated.
For example, suppose that a presynaptic spike fired at $\tilde{t}$
between $t_{k}$ and $t_{k+1}$. If one simply assigns it to be the end of time
step $t_{k+1}$, then the error of the spike-induced conductance change is
\begin{equation}
\frac{S}{N}[H(\sigma_{d}^{Q},\sigma_{r}^{Q},t-\tilde{t})-H(\sigma_{d}^{Q},\sigma_{r}^{Q},t-t_{k+1})]=O(t_{k+1}-\tilde{t})=O(\Delta t),Q=E,I.
\end{equation}
Therefore, the error with the magnitude of $\Delta t$ will be introduced when the system evolves to $t=t_{k+1}$. 

We now solve the above issue arising from the pulse-coupled
dynamics to achieve a second-order numerical accuracy.  First, we evolve the HH neural network without considering the feedforward and synaptic spikes during the time interval $[t_{k},t_{k+1}]$. Then, at time $t=t_{k+1}$, some neuron's voltage may be above the threshold, $i.e.$, generating a spike, say neuron $i$, if $V_i(t_k)<V^{\text{th}}$ and $V_i(t_{k+1}) \geq V^{\text{th}}$.
The spike time, say $\tau_{il}$, can be estimated
following the idea proposed in Ref. \cite{shelley2001efficient,Hansel1998}.
The neuron's membrane potential during the time interval can be approximated by a linear interpolation:
\begin{equation}
V_{i}(t)\approx V_{i,k}+\frac{V_{i,k+1}-V_{i,k}}{\Delta t}(t-t_{k}),\label{eq:linear V}
\end{equation}
and the spike time $\tau_{il}$ can be estimated by solving the equation:
\begin{equation}
V^{\textrm{th}}=V_{i,k}+\frac{V_{i,k+1}-V_{i,k}}{\Delta t}(\tau_{il}-t_{k}).\label{eq:spike time}
\end{equation}

Since there may be some neurons firing and some feedforward spikes emitting during the time interval and they will induce the conductance change, one should update the conductance by considering the effects of both the feedforward and synaptic spikes. 
When they are not considered, the conductance,
denoted by $\tilde{G}^{Q}$, is 
\begin{equation}
\tilde{G}_{j,k+1}^{E}=f\sum_{s_{jl}\leq t_{k}}H(\sigma_{d}^{E},\sigma_{r}^{E},t_{k+1}-s_{jl})+\frac{S}{N}\sum_{i}\sum_{\tau_{il}\leq t_{k}}H(\sigma_{d}^{E},\sigma_{r}^{E},t_{k+1}-\tau_{il}),
\end{equation}
\begin{equation}
\tilde{G}_{j,k+1}^{I}=\frac{S}{N}\sum_{i}\sum_{\tau_{il}\leq t_{k}}H(\sigma_{d}^{I},\sigma_{r}^{I},t_{k+1}-\tau_{il})
\end{equation}
and it should be recalibrated as 
\begin{equation}
G_{j,k+1}^{E}=\tilde{G}_{j,k+1}^{E}+f\sum_{t_{k}<s_{jl}\leq t_{k+1}}H(\sigma_{d}^{E},\sigma_{r}^{E},t_{k+1}-s_{jl})+\frac{S}{N}\sum_{i}\sum_{t_{k}<\tau_{il}\leq t_{k+1}}H(\sigma_{d}^{E},\sigma_{r}^{E},t_{k+1}-\tau_{il}),\label{eq:recalibrate GE}
\end{equation}
\begin{equation}
G_{j,k+1}^{I}=\tilde{G}_{j,k+1}^{I}+\frac{S}{N}\sum_{i}\sum_{t_{k}<\tau_{il}\leq t_{k+1}}H(\sigma_{d}^{I},\sigma_{r}^{I},t_{k+1}-\tau_{il}),\label{eq:recalibrate GI}
\end{equation}
for $j=1,2,...,N$. A detailed algorithm of the RK2 method is given in Algorithm 1.

\begin{algorithm}[H] 	
	\caption{RK2 algorithm}  	
	\KwIn{an initial time $t_k$ and feedforward input times $\{s_{il}\}$} 	 		 
	\KwOut{Solutions at time $t_{k+1}$}
	\For{$i = 1$ to $N$}  	
	{
		Solve the HH eqautions for the $i$th neuron without considering spike input using RK2 scheme.\\		 		
		\If{$V_i(t_k) < V^{\textrm{th}}$ and $V_i(t_{k+1}) \geq V^{\textrm{th}}$}		 	
		{
			The $i$th neuron spiked in $[t_k, t_{k+1}]$.\\	 		
			Estimate the spike time $\tau_{il}$ by Equation (\ref{eq:spike time}).			 
		} 		 
	} 
	Recalibrate the conductance by Equations (\ref{eq:recalibrate GE}) and (\ref{eq:recalibrate GI}).\\	
\end{algorithm} 

%

We show that the above algorithm can indeed achieve a second-order numerical accuracy as follows.
If there are no feedforward or synaptic spikes, then all the dependent
variables are infinitely differentiable and the RK2 method
can achieve an error of order $O(\Delta t^{2})$. For the
time step that contains feedforward or synaptic spikes, an error of order $O(\Delta t)$
is introduced in the conductance with the form of $G^{Q}-\tilde{G}^{Q},Q=E,I$.
Nevertheless, the dependent variables of $V,m,h$, and
$n$ can have an error of order $O(\Delta t^{2})$. The synaptic spike times are estimated by a linear interpolation and also have an error of order  $O(\Delta t^{2})$. After recalibration shown in Equations (\ref{eq:recalibrate GE})
and (\ref{eq:recalibrate GI}), the conductances can achieve numerical accuracy of second-order at the end of the time step.
Therefore, all the dependent variables $V,m,h,n,G^E$, and $G^I$ have an error of order $O(\Delta t^{2})$ (see below for verification of numerical results). 

\subsection{Exponential time differencing method}

Exponential time differencing method is proposed to solve the stiff problem in differential equations by decomposing the system into a linear stiff term and a nonlinear non-stiff term \cite{hochbruck1998exponential,cox2002exponential,kassam2005fourth,nie2008compact}. 
Following this idea, we propose the ETD schemes for HH Equations (\ref{eq: V of HH}) and (\ref{eq:mhn of HH}) below.
As illustrated in Algorithm 1, each neuron in the HH network is evolved independently and their conductances are recalibrated at the end of time step. Thus, one can first derive an ETD scheme for a single HH neuron and then consider the spike interactions among neurons, and obtain an ETD scheme for the numerical evolution of an HH neural network.

Consider the evolution of a single HH neuron from $t_k$ to $t_{k+1}$. We use $z_k$ to represent $z(t_k)$ for $z=V,m,h,n$ of this neuron and rewrite Equations (\ref{eq: V of HH}) and (\ref{eq:mhn of HH}) as 

\begin{equation}
\frac{dz}{dt}=c_{z}z+F_{z}, \,\,\, \text{ for }z=V,m,h,n,\label{eq:HH ETD}
\end{equation}
where
\begin{equation}
c_{V}=(-G_{Na}m_{k}^{3}h_{k}-G_{K}n_{k}^{4}-G_{L})/C,
\end{equation}

\begin{equation}
c_{z}=-\alpha_{z}(V_{k})-\beta_{z}(V_{k}), \,\,\, \text{ for }z=m,h,n,
\end{equation}
\begin{equation}
\begin{aligned}F_{V}(t,V,m,h,n) & =\left[-(V-V_{Na})G_{Na}m^{3}h-(V-V_{K})G_{K}n^{4}\right.\\
& \left.-(V-V_{L})G_{L}+I^{\textrm{input}}\right]/C-c_{V}V
\end{aligned}
\end{equation}
and 
\begin{equation}
F_{z}(t,V,m,h,n)=(1-z)\alpha_{z}(V)-z\beta_{z}(V)-c_{z}z,  \,\,\, \text{ for }z=m,h,n.
\end{equation}
Here, $F_z(t,V,m,h,n)$ is actually a function of $t,V$, and $z$ for $z=m,h,n$, but we write in this way for ease of illustration.  
Note that the linear coefficient
$c_{z}$ in Equation (\ref{eq:HH ETD}) is a constant value in the $k$th time
step $[t_{k},t_{k+1}]$ and is updated with respect to $k$.
Multiplying Equation (\ref{eq:HH ETD}) by an integrating factor $e^{-c_{z}t}$ and taking integral from $t_k$ to $t_{k+1}$, we obtain


\begin{equation} \label{eq: integral of ETD}
z_{k+1}=z_ke^{c_{z}\Delta t}+e^{c_{z}\Delta t}\int_{0}^{\Delta t}e^{-c_{z}\tau}F_{z}(t_k+\tau,V(t_k+\tau),m(t_k+\tau),h(t_k+\tau),n(t_k+\tau))d\tau
\end{equation}
for $z=V,m,h$ and $n$.

The essence of the ETD method is to derive proper approximations to
the above integration. We take a second-order ETD formulae with RK time stepping as follows \cite{cox2002exponential}.
Let
\begin{equation}
a_{z,k}=z_{k}e^{c_{z}\Delta t}+F_{z,k}(e^{c_{z}\Delta t}-1)/c_{z},\label{eq:ETD1}
\end{equation}
and approximate $F_z$ during the time interval $[t_k,t_{k+1}]$ by 
\begin{equation}
\begin{aligned} & F_{z}(t_{k}+\tau,V(t_{k}+\tau),m(t_{k}+\tau),h(t_{k}+\tau),n(t_{k}+\tau)\\
& =F_{z,k}+\tau(F_{z}(t_{k+1},a_{V,k},a_{m,k},a_{h,k},a_{n,k})-F_{z,k})/\Delta t+O(\Delta t^{2}),
\end{aligned}
\end{equation} 
for $z=V,m,h$, and $n$, where $F_{z,k}$ represents $F_{z}(t_{k},V_k,m_k,h_k,n_k)$.
Substituting the above approximation into Equation (\ref{eq: integral of ETD}) yields the ETD2 scheme which is given by

\begin{equation}
z_{k+1}=a_{z,k}+[F_{z}(t_{k+1},a_{V,k},a_{m,k},a_{h,k},a_{n,k})-F_{z,k}](e^{c_{z}\Delta t}-1-c_{z}\Delta t)/c_{z}^{2}\Delta t,\label{eq:ETD2}
\end{equation}
for $z=V,m,h$, and $n$. 
The procedure of the ETD2 algorithm for an HH neural network is similar to that of the RK2 algorithm given in Algorithm 1, but the RK2 scheme is replaced by the ETD2 scheme in Equation (\ref{eq:ETD2}).

\subsection{Adaptive Exponential time differencing method}
The ETD2 method can indeed use a large time step to improve computational efficiency, but we find that it will introduce relatively large error in the trajectories of neurons' membrane potentials and even lead to the missing of action potentials (see below for numerical results). In addition, the number of the missed action potentials in the ETD2 method can grow with the increase of time steps compared with the RK2 method using a small time step.  
Thus, it is important to design an efficient but also reliable ETD method to solve this issue.

\begin{figure}[H]
	\begin{centering}
		\includegraphics[width=1\textwidth]{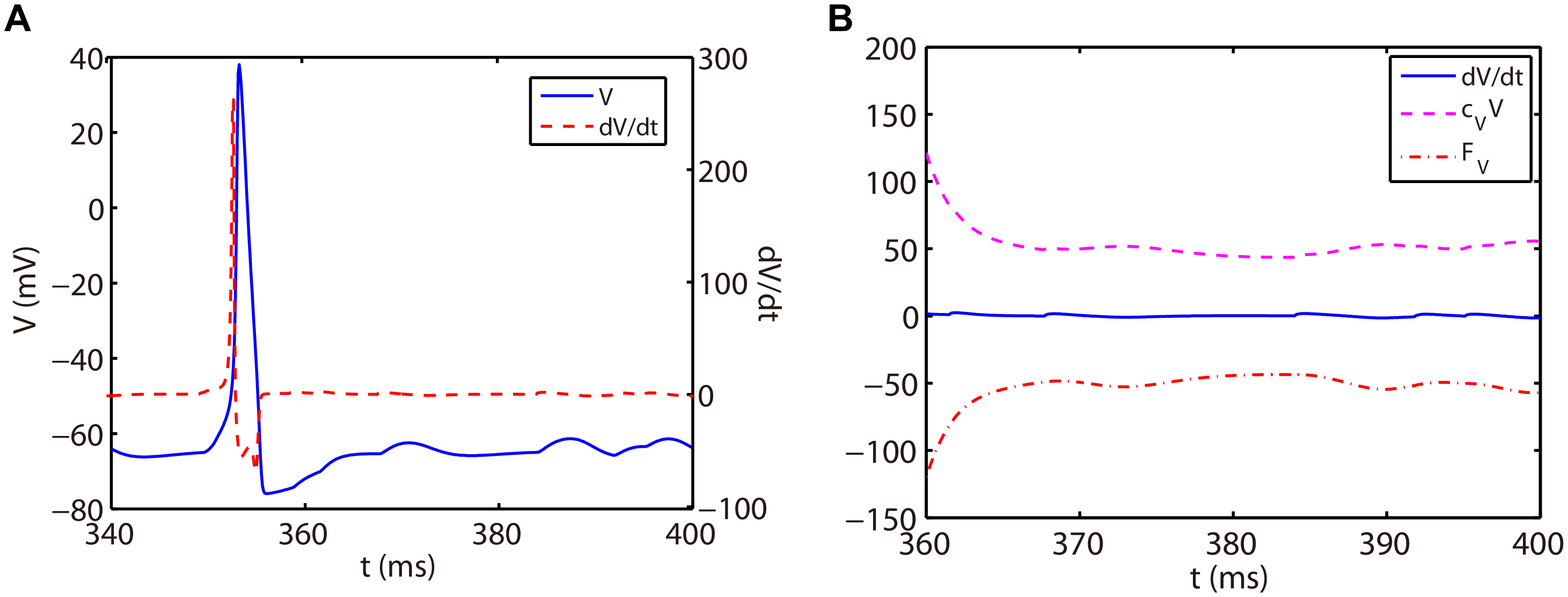}
		\par\end{centering}
	\caption{(A) Trajectory of voltage (blue solid curve) and the slope of voltage (red dashed curve) for a single HH neuron.  (B) Trajectory of the slope of voltage (blue solid curve), linear part $c_VV$ (magenta dashed curve), and nonlinear part $F_V$ (red dash-dotted curve) in Equation (\ref{eq:HH ETD}) for the non-spike period. 
	}\label{fig:ETD fail}
\end{figure}

As shown in Figure \ref{fig:ETD fail}A, the slope of voltage has a very large value when the neuron generates an action potential (spike period) and quickly reduces to a value around zero in the non-spike period until the next spike time. Therefore, the stiffness of HH equations is quite different between spike and non-spike periods. In the non-spike period, the slope of voltage is 
almost zero, while the linear and nonlinear part in Equation (\ref{eq:HH ETD}) have a much larger absolute value and nearly cancel each other out as shown in Figure \ref{fig:ETD fail}B. Therefore, the decomposition in Equation (\ref{eq:HH ETD}) may not be appropriate in the non-spike period since both the linear and nonlinear parts become stiff while the summation of them is indeed non-stiff. Based on this, we propose a different decomposition in the non-spike period from that in the spike period: taking $c_z=0$ and the whole right hand side of Equation (\ref{eq:HH ETD}) as the nonlinear part. For such a decomposition, the ETD2 scheme reduces to the RK2 scheme in the non-spike period.

\begin{figure}[H]
	\begin{centering}
		\includegraphics[width=0.5\textwidth]{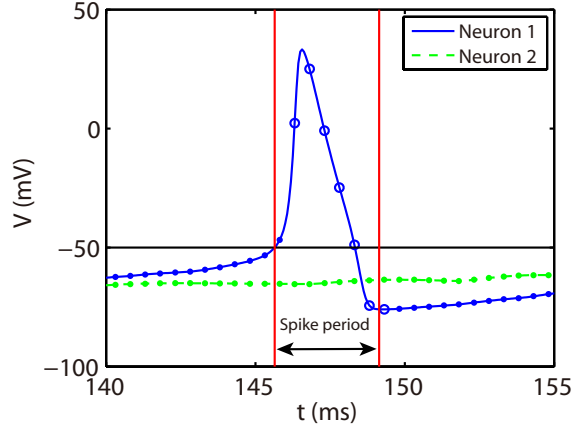}
		\par\end{centering}
	\caption{Illustration of the AETD2 method. After neuron 1 fires a spike, we use the ETD2 scheme to evolve the HH equations for neuron 1 during the spike period indicated by the red vertical lines, while the HH equations for neuron 2 is evolved using RK2 scheme since neuron 2 is in the non-spike period. The starting point of the spike period can be determined as the spike time and it lasts for the following about 3.5 ms. The circles and dots indicate the time nodes where we use the ETD2 and RK2 schemes, respectively.
		\label{fig:Illustration-AETD2}}
\end{figure}

Based on the above observation, we give our AETD2 method for HH neural network as following: each neuron is evolved using ETD2 scheme if it is in the spike period and use the reduced ETD2 scheme, the RK2 scheme, otherwise, as shown in Figure \ref{fig:Illustration-AETD2}. The starting point of the spike period can be determined as the spike time and the interval of spike period can be determined as 3.5 ms which is sufficient large to cover the highly stiff region of the spike. Detailed AETD2 algorithm is given in Algorithm 2.

\begin{algorithm}[H] 	    
	\caption{AETD2 algorithm}  	 	
	\KwIn{an initial time $t_k$,  feedforward input times $\{s_{il}\}$} 	 			
	\KwOut{Solutions at time $t_{k+1}$}
	
	\For{$i = 1$ to $N$}  	
	{
		Solve the HH equaions for the $i$th neuron without considering spike input:\\
		
		\uIf{The $i$th neuron is inside spike period}
		{
			use ETD2 scheme
		}	 
		\Else	
		{
			use RK2 scheme
		}		 				 
		
		\If{$V_i(t_k) < V^{\textrm{th}}$ and $V_i(t_{k+1}) \geq V^{\textrm{th}}$}		 	
		{
			The $i$th neuron spiked in  $[t_k, t_{k+1}]$.\\	 		
			Estimate the spike time $\tau_{il}$ by  Equation (\ref{eq:spike time}).			 
		} 		 
	} 
	Recalibrate the conductance by Equations (\ref{eq:recalibrate GE}) and (\ref{eq:recalibrate GI}).			
\end{algorithm}

%

\section{RESULTS} \label{sec: Numerical result}

We consider an all-to-all connected network of 80 excitatory and 20
inhibitory neurons driven by Poisson feedforward input. For the ease of illustration,
we choose the Poisson input strength $f=0.06$ $\textrm{mS\ensuremath{\cdot}cm}^{-2}$ and input rate
$\nu=300$ Hz, and the coupling strength between neurons are chosen as $S=0.2$ $\textrm{mS\ensuremath{\cdot}cm}^{-2}$
throughout this work, unless indicated otherwise. 
However, our algorithm can be applied to HH neural networks under a variety of dynamical regimes.

\begin{figure}[H]
	\begin{centering}
		\includegraphics[width=1\textwidth]{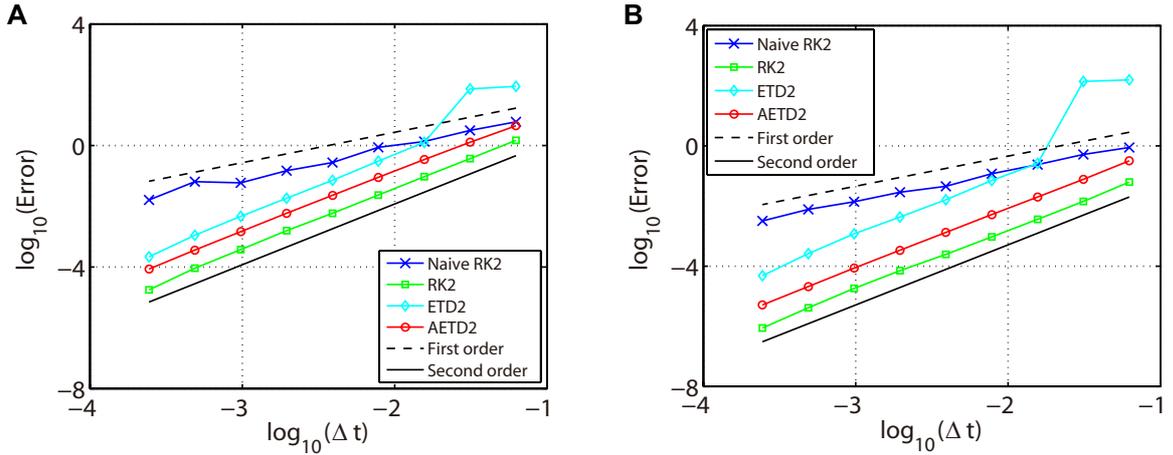}
		\par\end{centering}
	\caption{Errors of membrane potentials (A) and the last spike time of each neuron (B) in the all-to-all connected network when it is evolved using various time steps. Blue crosses are naive RK2  method without performing the linear interpolation for the estimate of the spike times. Green
		squares are RK2 method, cyan diamonds
		are ETD2 method, and red circles are AETD2 method. The last three methods all perform the linear 
		interpolation to estimate the spike times. The dashed line and the solid line indicate the numerical convergence of the first-order and the second-order, respectively. The total run time $T=2000$ ms. \label{fig:convergence test}}
\end{figure}

First, we verify the second-order numerical accuracy by performing
convergence tests. A high precision solution is obtained by using
RK2 method with a sufficiently small time step $\Delta t=1\times10^{-6}$
ms and is denoted by a superscript "high". It is compared with the solutions computed by the RK2, ETD2, and AETD2
methods with various values of larger time steps $\Delta t=2^{-4},2^{-5},...,2^{-12}$
ms which is denoted by a superscript "$\Delta t$". Errors of membrane potentials at final run time $T=2000$ ms and the
last spike time of each neuron are computed:

\begin{equation}
\text{Error}_{V}=\sqrt{\sum_{i}(V_{i}^{(\Delta t)}(T)-V_{i}^{(\text{High})}(T))^{2}},
\end{equation}

\begin{equation}
\text{Error}_{\tau}=\sqrt{\sum_{i}(\tau_{il^{*}}^{(\Delta t)}-\tau_{il^{*}}^{(\text{High})})^{2}},
\end{equation}
where $\tau_{il^{*}}$ indicates the last spike time of the $i$th neuron during the run time interval.
As shown in Figure \ref{fig:convergence test}, if one naively
assigns the end of time step as the spike times in the RK2 method, the numerical accuracy of the membrane potentials and spike times can only be of the first-order. 
In contrast, if one determines the spike times by linear interpolation and
recalibrate the conductances accordingly, all the RK2, ETD2, and
AETD2 methods can achieve a second-order numerical accuracy. 
In addition, we find that the ETD2 method has much larger error compared with the RK2 and AETD2 methods using the same time step as shown in Figure \ref{fig:convergence test}. When using a time step larger than $\Delta t=2^{-6}=0.0156$ ms, the ETD2 method performs even worse than the naive RK2 method.  The underlying reason is that the HH equations are almost non-stiff in the non-spike period, but the decomposition in Equation (\ref{eq:HH ETD}) induces a relatively large stiffness for the nonlinear term as discussed previously.

\begin{figure}[H]
	\begin{centering}
		\includegraphics[width=1\textwidth]{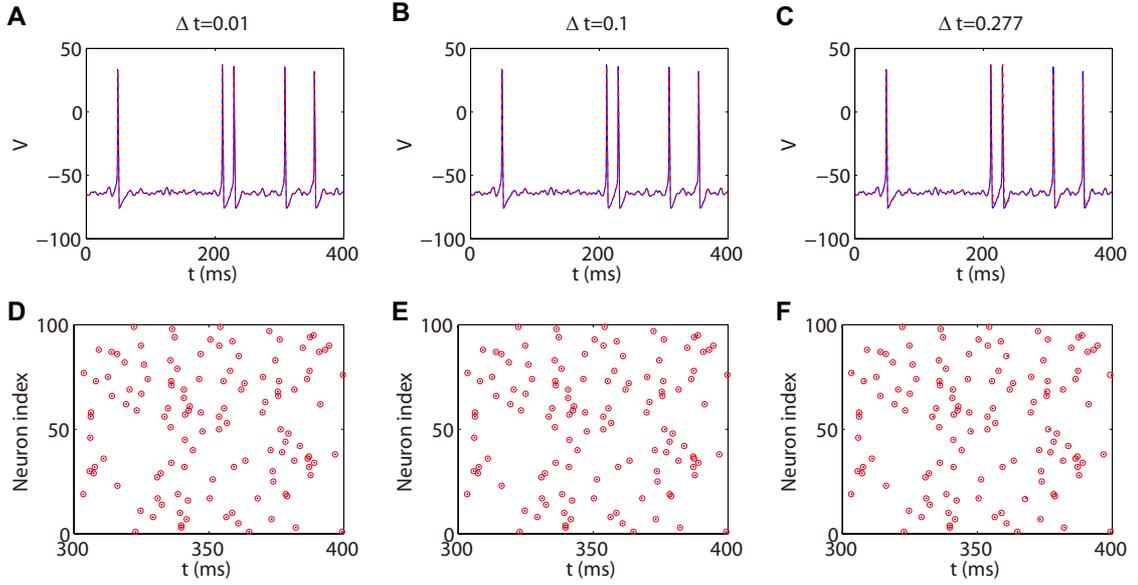}
		\par\end{centering}
	\caption{(Top panel): Traces of membrane potential of an HH neuron in the all-to-all connected network.
		(Bottom panel): Raster plots of the network spikes. The blue solid
		curves and dots indicate the results by the RK2 method with time step $\Delta t=0.01$ms, while
		the red dashed curves and circles indicate the results by the AETD2 method. The
		time steps for the AETD2 method are $\Delta t=0.01,0.1,0.277$ ms
		for (A, D), (B, E) and (C, F), respectively. \label{fig: V and raster}}
\end{figure}

We next discuss the numerical performance of our AETD2 method and compare it with other different numerical methods.
As shown in the top panel of Figure \ref{fig: V and raster}, the
AETD2 method with large time steps (maximum time step $\Delta t=0.277$
ms) can obtain the same high accuracy in membrane potentials as
the RK2 method using a very small time step $\Delta t=0.01$ ms.  The bottom
panel of Figure \ref{fig: V and raster} shows the raster plots (neuron
index versus its spike time) of the spike events in the network.  
It can be seen that the spike times are well captured by the AETD2 method with large time steps.

\begin{figure}[H]
	\centering{}\includegraphics[width=1\textwidth]{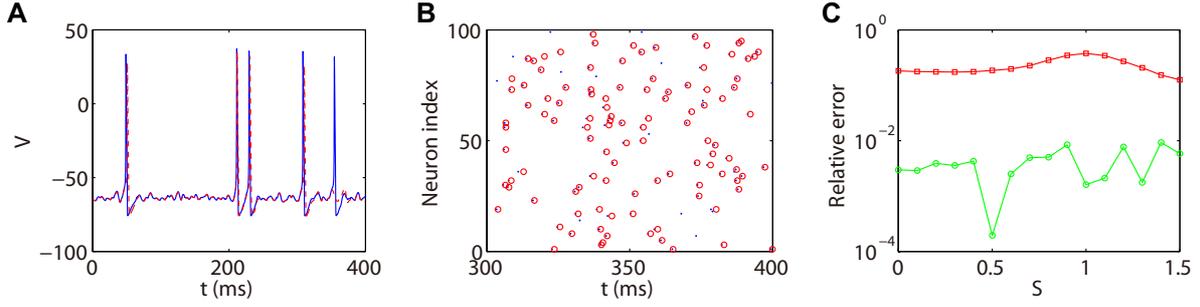}
	\caption{(A) Voltage trace of the same HH neuron used in Figure \ref{fig: V and raster}. (B) Raster plot of
		the network spikes. The blue solid curve and dots indicate the results by the RK2
		method with time step $\Delta t=0.01$ms while the red dashed curve and circles
		indicate the results by the ETD2 method with time step $\Delta t=0.277$ ms. The
		coupling strength is $S=0.2$ $\textrm{mS\ensuremath{\cdot}cm}^{-2}$.
		(C) Relative error in the mean firing rates between the ETD2 (AETD2) and the RK2 methods for different choice of the coupling
		strength. The red squares indicate
		ETD2 method and the green circles indicate AETD2 method. Both the ETD2 and AETD2 methods use time step $\Delta t=0.277$ ms. The benchmark mean firing rate is computed by the RK2 method with a very small time step $\Delta t=1\times10^{-6}$
		ms.  \label{fig:std ETD2}}
\end{figure}

The ETD2 method is proved to be unconditionally stable for the HH system in Ref. \cite{borgers2013exponential}. Hence, it can use a much larger time step compared with AETD2 method in principle. However, as shown in Figure \ref{fig:std ETD2}, the ETD2 method is highly inaccurate when the time step $\Delta t=0.277$ ms is used (the maximum time step in AETD2 method). 
There are some spikes missing as shown in Figure \ref{fig:std ETD2}A and Figure \ref{fig:std ETD2}B, in contrast, the AETD2 method does not encounter such issues when using the same time step. 
Figure \ref{fig:std ETD2}C shows the relative error in the mean firing rate (the average
number of synaptic spikes per unit time) between the RK2 and ETD2 (AETD2) methods over different values of coupling strength. It can be seen that the ETD2 method can achieve only one digit of numerical accuracy while the AETD2 method can robustly achieve more than two digits of numerical accuracy when the time step $\Delta t=0.277$ ms is used in both methods. Therefore, the ETD2 method has worse numerical accuracy in terms of voltage traces and firing rates compared with the AETD2 method. 

To demonstrate the efficiency of our AETD2 method, we compare the
simulation time that RK2, ETD2, and AETD2 methods take for a common total
run time. We simulate the all-to-all connected network by RK2, ETD2, and AETD2 methods on Visual Studio using an Intel i7 2.6 GHz processor, and the simulation time and numerical accuracy of mean firing rate are given in Table \ref{tab:simulation time_rate}.
The AETD2 method can achieve over an order of magnitude of speedup compared with the RK2 method while achieving the same high accuracy in terms of the mean firing rate. 

\begin{table}[H]
	\begin{centering}
		\begin{tabular}{|c|c|cc|c|cc|c|cc|}
			\hline 
			& \multicolumn{3}{c|}{RK2} & \multicolumn{3}{c|}{ETD2} & \multicolumn{3}{c|}{AETD2}\tabularnewline
			\hline 
			$\Delta t$ (ms) & CPU  & \multicolumn{2}{c|}{Relative error } & CPU & \multicolumn{2}{c|}{Relative error} & CPU & \multicolumn{2}{c|}{Relative error}\tabularnewline
			\hline 
			0.005 & 60.56 s & 0  & (13.61 Hz) & 60.07 s & 0  & (13.61 Hz) & 60.82 s & 0  & (13.61 Hz)\tabularnewline
			\hline 
			0.01 & 30.22 s & 0  & (13.61 Hz) & 30.05 s & 0  & (13.61 Hz) & 30.55 s & 0  & (13.61 Hz)\tabularnewline
			\hline 
			0.02 & 14.99 s & 0  & (13.61 Hz) & 15.03 s & 0.074 \%  & (13.60 Hz) & 15.30 s & 0  & (13.61 Hz)\tabularnewline
			\hline 
			0.05 & {*}{*}{*} & {*}{*}{*} & {*}{*}{*} & 5.95 s & 0.66 \%  & (13.52 Hz) & 6.16 s & 0.074 \%  & (13.62 Hz)\tabularnewline
			\hline 
			0.1 & {*}{*}{*} & {*}{*}{*} & {*}{*}{*} & 2.94 s & 2.65 \%  & (13.25 Hz) & 3.11 s & 0.074 \%  & (13.62 Hz)\tabularnewline
			\hline 
			0.2 & {*}{*}{*} & {*}{*}{*} & {*}{*}{*} & 1.48 s & 9.99 \%  & (12.25 Hz) & 1.57 s & 0.15 \%  & (13.63 Hz)\tabularnewline
			\hline 
			0.277 & {*}{*}{*} & {*}{*}{*} & {*}{*}{*} & 1.09 s & 12.56 \%  & (11.29 Hz) & 1.15 s & 0.59 \%  & (13.69 Hz)\tabularnewline
			\hline 
			0.5 & {*}{*}{*} & {*}{*}{*} & {*}{*}{*} & 0.57 s & 41.59 \%  & (7.95 Hz) & {*}{*}{*} & {*}{*}{*} & {*}{*}{*}\tabularnewline
			\hline 
			1 & {*}{*}{*} & {*}{*}{*} & {*}{*}{*} & 0.29 s & 87.07 \%  & (1.76 Hz) & {*}{*}{*} & {*}{*}{*} & {*}{*}{*}\tabularnewline
			\hline 
		\end{tabular}
		\par\end{centering}
	\caption{Simulation of the all-to-all connected network with a total run time
		$T=10$ seconds. The simulation time is measured in seconds. The relative
		error in the mean firing rate between each method using different
		time steps and the RK2 method using a very small time step $\Delta t=1\times10^{-6}$ ms is measured
		in percentage and the mean firing rate is measured in Hz given inside the parentheses. Asterisks indicate overflow errors. \label{tab:simulation time_rate}}
\end{table}

\begin{figure}[H]
	\centering{}\includegraphics[width=1.0\textwidth]{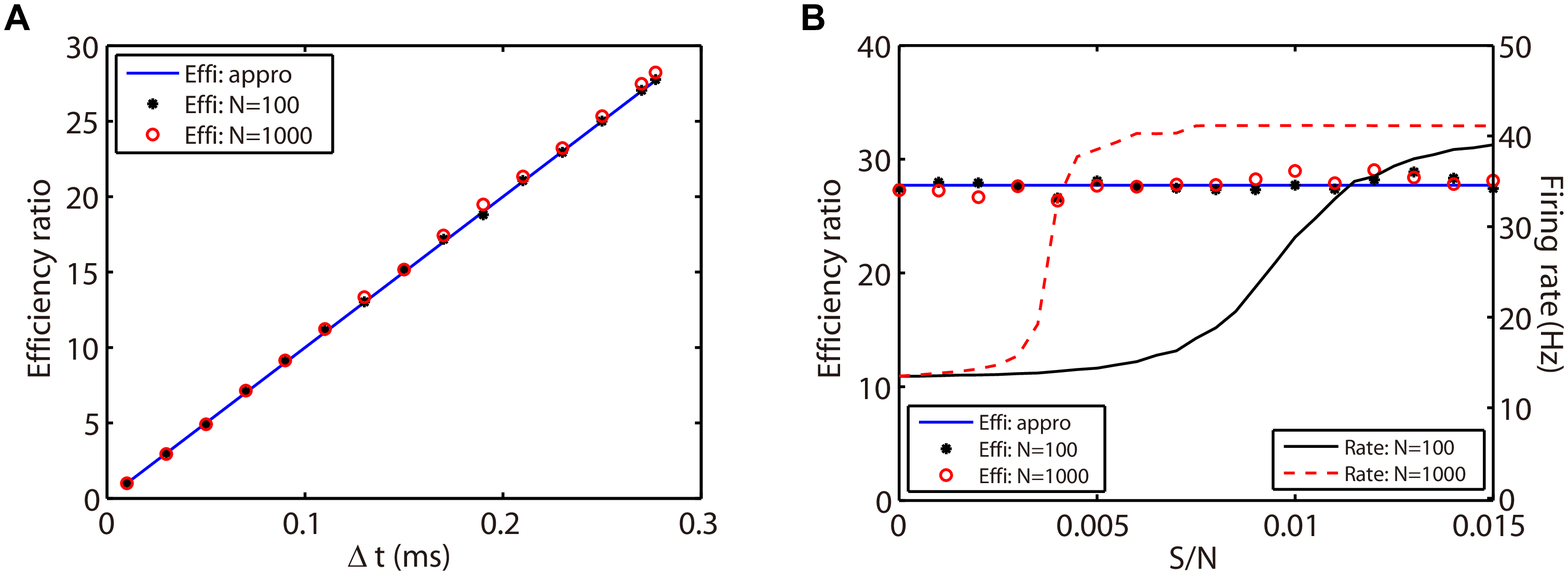}
	\caption{Efficiency ratio of the AETD2 method when evolving the HH neural network using various time steps (A) and coupling strength
		(B). In (A) and (B), the blue lines are efficiency ratio measured by the approximation in Equation (\ref{eq:Effi dt}), while the black stars and red circles are the efficiency ratio measured by
		the ratio of simulation times between the RK2 and AETD2 methods. The black stars represent the results for the all-to-all
		connected HH neural network of 80 excitatory and 20 inhibitory neurons, while
		the red circles represent the results for an HH neural network of 800 excitatory and 200 inhibitory
		neurons randomly connected with probability 25\%. The black solid and red dashed curves in (B) are the mean firing rates in the smaller network of 100 neurons and larger network of 1000 neurons, respectively.
		The coupling strength
		in (A) is $S/N=0.002$ $\textrm{mS\ensuremath{\cdot}cm}^{-2}$ and the
		time step for AETD2 method in (B) is $\Delta t=0.277$ ms. The time
		step for RK2 method is $0.01$ ms and total
		run time is $T=50$ seconds in both (A) and (B). \label{fig:efficiency}}
\end{figure}

In addition, we define the efficiency ratio of the AETD2 method over the RK2 method as
\begin{equation}
E=\frac{T_{\text{RK2}}}{T_{\text{AETD2}}}
\end{equation}
where $T_{\text{RK2}}$ and $T_{\text{AETD2}}$ indicate the simulation times of the RK2 and AETD2 methods, respectively, for the HH neural network to evolve the run time $T$.
Note that the RK2 and ETD2 methods take almost the same simulation time when using the same small time step as shown in Table \ref{tab:simulation time_rate}.
Thus, the above efficiency ratio can be approximated by the ratio of the total number of time steps each method requires as
\begin{equation}
E\approx\frac{T/\Delta t_{\text{RK2}}}{T/\Delta t_{\text{AETD2}}}=\frac{\Delta t_{\text{AETD2}}}{\Delta t_{\text{RK2}}},\label{eq:Effi dt}
\end{equation}
where $\Delta t_{\text{RK2}}$ and $\Delta t_{\text{AETD2}}$ indicate the time steps used in the RK2 and
AETD2 methods, respectively. 
To demonstrate that the above efficiency ratio is independent of the network connectivity and size, we
evolve the all-to-all connected network of 80 excitatory and 20 inhibitory neurons and a randomly connected network 
of 800 excitatory and 200 inhibitory neurons. As shown in Figure \ref{fig:efficiency}A,
the efficiency ratio approximated by Equation (\ref{eq:Effi dt}) agrees well
with the one measured by the ratio of simulation time between the RK2 and
AETD2 methods in both two networks. These two networks are further evolved with a variety choice of coupling strength to cover a wide range of different dynamical regimes as shown in Figure \ref{fig:efficiency}B. Nevertheless, the efficiency ratio measured by the ratio of simulation time is still consistent with the one approximated by Equation (\ref{eq:Effi dt}) in both networks.
Hence, the efficiency ratio of the AETD2 method relies
on only the number of evolved time steps and appears to be independent of parameter choice of coupling strength, connectivity structure and network size. 

\section{Discussion}

We have presented an adaptive second-order ETD method to evolve the pulse-coupled
HH neural network. 
Our AETD2 method can solve the stiff problem in the HH equations
when an HH neuron generates an action potential (spike period). It can use a large time step to raise
computational efficiency while accurately capturing dynamical properties of HH neurons such as
the trace of membrane potentials, spike times of each neuron and the mean firing rate. We
point out that our AETD2 method can robustly enlarge time steps and
raise computational efficiency over one order of magnitude compared with the 
RK2 method. This high efficiency seems to be independent of parameter choice of 
connectivity structure, dynamical regimes or network size.

We should point out that our decomposition of HH equations in Equation 
(\ref{eq:HH ETD}) is an example to apply the ETD method. Other forms
of decomposition can be similarly derived under the framework of the ETD method
in future. Here, we use the ETD2 scheme derived by 
approximating the integration in Equation (\ref{eq: integral of ETD}) with RK time stepping. 
Other forms of numerical schemes can also be used to approximate the integration. For example, one can use a liner interpolation to approximate the nonlinear part in Equation (\ref{eq:HH ETD}) to obtain another form of ETD2 scheme. The high efficiency and numerical accuracy
can be obtained in our additional numerical experiments. 

In this work, the numerical accuracy of our AETD2 method is second-order. In some situations,
high accurate traces of membrane potentials may be required, especially
the accurate shape of action potentials  \cite{traub2001gap,kopell2004chemical}.
Therefore, one future work may be the design of the fourth-order ETD method. As illustrated
above, due to the discontinuity arising from the pulse-coupled dynamics, an even more
careful recalibration needs to be designed to achieve a fourth-order numerical accuracy. 

Finally, we point out that our AETD2 method can be easily extended
to networks of other HH type neurons
\cite{pospischil2008minimal}. And our AETD2 method can also robustly
achieve high numerical accuracy and  efficiency.  
In addition, our method is naturally a parallel algorithm which can be applied to simulations of large-scale neural network dynamics.

\section*{Funding}
This work was supported by National Science Foundation of China with Grant No. 11671259, 11722107,
91630208 and SJTU-UM Collaborative Research Program (D.Z.) 

\section*{Acknowledgments}	
We thank the Student Innovation Center at Shanghai Jiao Tong University, and we dedicate this paper to our late  professor David Cai.

\begin{small}
\bibliographystyle{plain}
\bibliography{ETDrefer}
\end{small}

\end{document}